\documentstyle[prb,aps,twocolumn,graphicx]{revtex}

\begin{document}


\title{Magnetization reversal and spin dynamics exchange in biased  F/AF bilayers probed with complex permeability spectra }
\author{David Spenato, Jamal Ben Youssef and Henri Le Gall  }
\address{Laboratoire de Magn\'etisme de Bretagne, CNRS/UPRESA 6135, Universit\'e, 6 avenue Le Gorgeu, 29285 Brest,
France}
\wideabs{ \maketitle
\begin{abstract}
The spin dynamics of the ferromagnetic pinned layer of
ferro-antiferromagnetic coupled  NiFe/MnNi bilayers is
investigated in a broad frequency range (30 MHz-6 GHz). A
phenomenological model based on the Landau-Lifshitz equation for
the complex permeability of the F/AF bilayer is proposed. The
experimental results are compared to theoretical predictions. We
show that the resonance frequencies, measured during the
magnetization, are likewise hysteretic.
\end{abstract}

\pacs{} }


\section{Introduction}

Despite many theoretical and experimental studies in the last
decade on and old \cite{meik} phenomenon known as "exchange
biasing", a full understanding of the exchange anisotropy at the
interface between antiferromagnets (AF) and ferromagnets (F) is
not clearly understood. One of the interesting observations is
that irreversible measurements (such as hysteresis loop) and
reversible techniques which imply small perturbations of the
magnetization around equilibrium (AC susceptibility \cite{susc},
ferromagnetic resonance \cite{fmr}, Brillouin light scattering
\cite{bls}, anisotropic magnetoresistance \cite{amr}) seem to lead
to different values of the exchange field. The second great
challenge is to understand the mechanism by which the
magnetization reverses in such systems. The aim of this paper is
to probe by means of complex permeability spectra the exchange
anisotropy of NiFe/MnNi bilayers and to study the magnetization
reversal in such bilayers.

\section{EXPERIMENTAL PROCEDURE}

 Substrate/$Ni_{81}Fe_{19}$ ($t_{F}$)$/Mn_{50}Ni_{50}$ ($t_{AF}$)
 bilayers were grown on Corning Glass substrate. After deposition,
samples were annealed in a magnetic field of 1000 Oe, aligned with
the easy axis of the (F) film, at $300^o$C for 5 hours to induce
the exchange field. Detail of the sample preparation have been
published elsewhere\cite{spen_1}. The magnetic properties such as
the saturation magnetization $M_{s}$ and the coercivity $H_{c}$
were obtained from magnetization loops (VSM) measured at room
temperature. The coercivity of the biased NiFe layer was defined
by the half of the shifted M-H loop width. The complex frequency
spectra (CPS) of the bilayers were measured from 30 MHz to 6 GHz
using a broad band method based on the measurement, by a network
analyser, of the reflection coefficient $S_{11}$ of a single turn
coil loaded by the film under test \cite{monos1}. Because of the
topography of the applied ac field ($h_{ac}$) in the coil, the
permeability can be measured for different orientations of the
exciting field in relation to the in-plane anisotropy. Amoung the
prepared samples, we have chosen the ones with low values of the
exchange biasing and coercive field essential for dtecting a
signal in the CPS measurements, all the results presented in this
paper have been made on the same sample.

\section{RESULTS AND DISCUSSION}

\begin{figure}
\begin{center}
\includegraphics[width=7.5 cm]{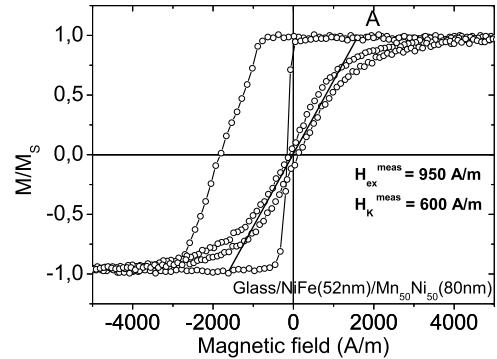}\caption{Typical easy- and hard-axis hysteresis loops for a
Glass/NiFe 52nm/$Mn_{50}Ni_{50}$ 80nm bilayer.}
 \label{figure1}
\end{center}
\end{figure}

\begin{figure}
\begin{center}
\includegraphics[width=7 cm]{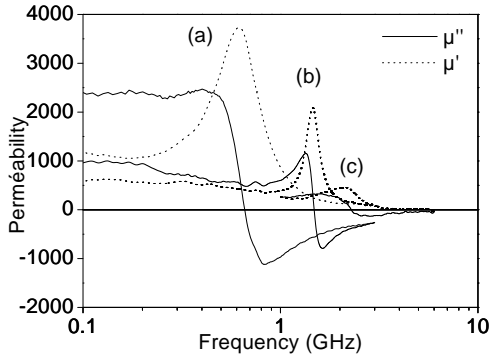} \caption{Complex
permeability spectra of a NiFe 52nm/$Mn_{50}Ni_{50}$ 80nm bilayer.
(a) as-grown; (c) annealing at 300°C for 5 hours ($H_{S}=950 A/m$;
(b) as-grown with an applied static field of 1550 A/m }
\label{figure2}
\end{center}
\end{figure}

We consider a F layer, with an uniaxial anisotropy field H$_{k}$
along the easy axis, submitted to an unidirectionnal exchange
field H$_{ex}$, induced by the exchange coupling along the easy
axis, and to an applied ac field $h_{ac}$ applied perpendicular to
the easy axis. The initial susceptibility measured along the ac
field is given by \cite{xi} $\chi=M_{S}/(H_{ex}+H_{k})$.
$\chi/M_{S}$ is the initial slope of the hysteresis loop when
measured perpendicular to the easy axis. $H_{ex}^{meas}$ is
determined by the shift of the center of the magnetization loop.
Following Xi and White \cite{xi2},the susceptibility may be
obtained from the hard-axis hysteresis loop by drawing a line
tangent to the loop trough the origin intersecting the asymptotic
limits of $M_S$ (point A). $H_{k}^{meas}$ is extracted from the
M-H loops measured perpendicular to the easy  axis (previous
equation). Fig. \ref{figure1} shows easy- and hard-axis loops for
our bilayers.
The extract values of $H_{ex}^{meas}$ and
$H_{ex}^{meas}$ are mentioned on this figure. In our samples the
forward loop shows a slope, as if the full magnetization  take
place progressively, while the reverse loop is very very square
which may be the consequence of either coherent magnetization flip
(such as in monodomain particles) or the presence of high mobility
domain walls. We have measured CPS of the bilayers before and
after annealing for the exciting field applied perpendicular to
the easy axis. An example of the measurement on a NiFe/MnNi
bilayer is presented in Fig.\ref{figure2}.

 The as deposited state of the MnNi
is non magnetic fcc structure and there is no magnetic interaction
between the NiFe and the MnNi layer. The permeability spectra are
typical of damping by spin rotation processes in a NiFe layer
Fig.\ref{figure2}(a) with a resonance frequency of $\mu''$ at
about 700 MHz\cite{oliv}. In order to achieve an antiferromagnetic
tetragonal (fct) state, a high temperature annealing was
performed. For the annealed samples, the level of the real part of
the permeability $\mu'(0)$ at low frequency decreases and the roll
off frequency increases fig.\ref{figure2}(c). We can also observe
that the imaginary part of the complex permeability $\mu''$ shows
a lower resonance peak, a higher resonance frequency $f_{res}$(2.5
GHz) and a wider resonance peak as the exchange field increases.
We have submitted to a static magnetic field whose amplitude is
equivalent to ($H_{k}^{meas}$ +$H_{ex}^{meas}$ = 1550 A/m)) the
as-grown sample and measured the CPS, the results are described on
fig.\ref{figure2}(b). It is clear that such static field is not
sufficient to reach the high resonance frequency of the annealed
bilayer and the spectra are still very resonant.

A general description of the spin dynamics is often made using the
Landau Lifshitz (LL) theory, a useful representation is the
Gilbert form of the equation of motion:

\begin{equation}
\frac{d\mathbf{M}}{dt}=\mu_{0}\gamma(\mathbf{M}\land\mathbf{H})-\frac{\alpha\mu_{0}\gamma}{M}(\mathbf{M}\land(\mathbf{M}\land\mathbf{H}))
\end{equation}

In a previous paper \cite{dave2}, we have presented an analytic
calculation of the frequency dependent complex permeability
tensor of a thin ferromagnetic film with uniaxial in-plane
anisotropy, submitted to an external exciting field using the
Landau-Lifshitz (LL) theory \cite{landau}. Using this
calculation, we have obtained the components of the complex
permeability tensor which are a function of $M_{s}$, the total
effective field $H_{eff}$, the frequency f of the exciting field
and the phenomenological damping constant $\alpha$. The
theoretical value of $\mu'(0)$ at low frequency is found to be
1+($M_{s}/H_{eff}$) and, as observed, the decrease of the
saturation magnetization and the enhancement of the effective
field ($H_{k}^{mes}+H_{ex}^{mes}$) lead to a reduction of the
level of $\mu'(0)$. The resonance frequency $f_{res}$ is found to
be:
\begin{equation}
 \frac{1}{2\pi}\times\gamma(H_{eff}(H_{eff}+M_{s}))^{1/2}.
 \end{equation}

\begin{figure}
\begin{center}
\includegraphics[width=7.5 cm]{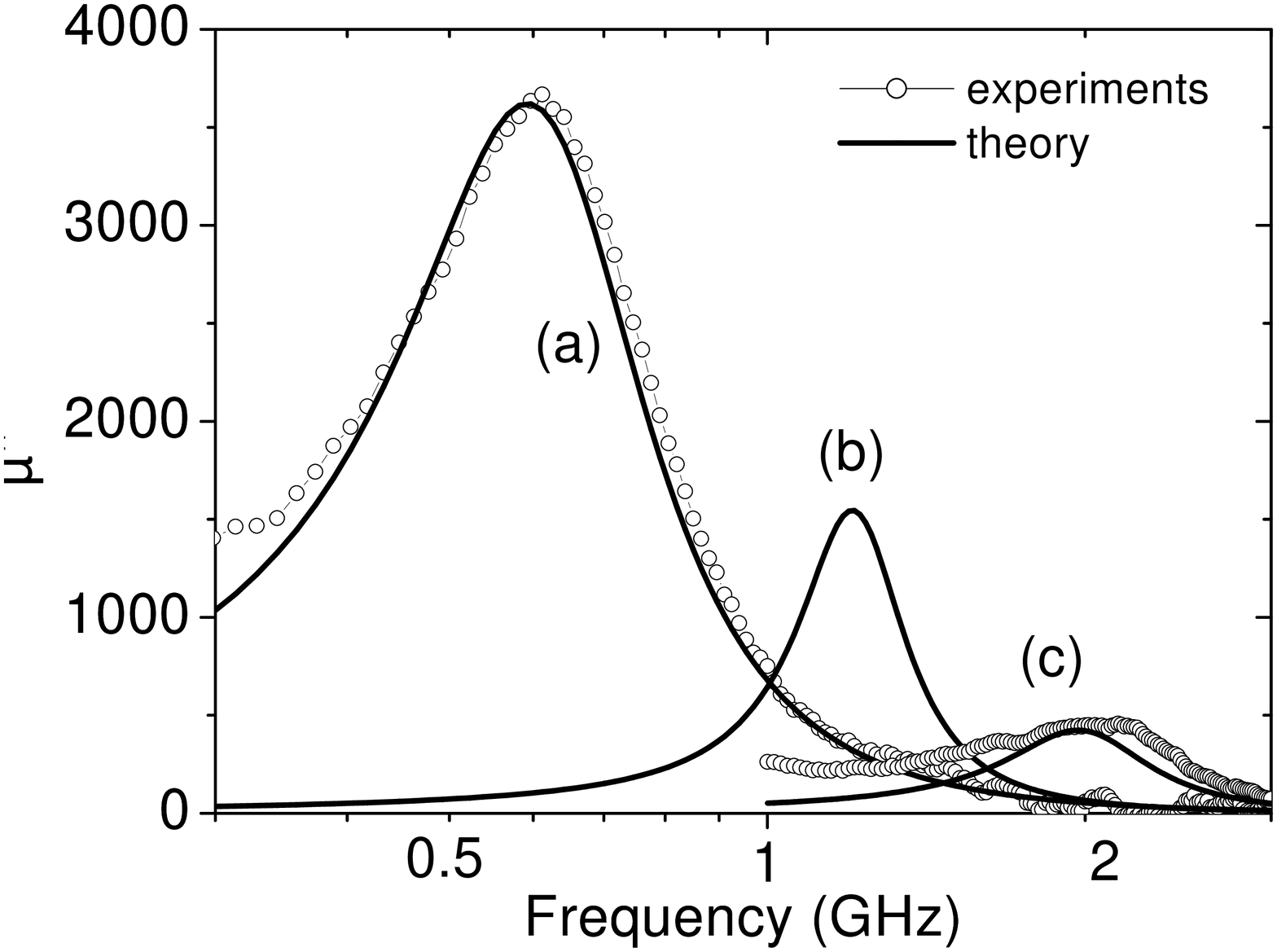} \caption{Measured and
calculated $\mu''$ spectra of an NiFe/MnNi bilayer when the
exciting field is applied along the hard axis; Solid line (a)
simulated curve, (b) fitted curve} \label{figure3}
\end{center}
\end{figure}

In our samples the enhancement of $H_{eff}^{mes}$ is prevalent and
lead to the enhancement of $f_{res}$. Fig.\ref{figure3} shows the
comparison between theoretical and experimental complex
permeability spectra when the exciting field is applied
perpendicular to the easy axis. In a first step of calculations,
the values the effective field ($H_{k}^{meas}+H_{ex}^{meas}$) and
$M_{s}^{meas}$  are taken from static measurements and the value
of the damping parameter is fitted. For the as-grown sample
$(H_{ex}^{meas}=0, H_{K}^{meas}=360 A/m, M_{S}^{meas}=800 kA/m )$
experimental results are in good agreement with the theoretical
prediction as observed in fig. \ref{figure3} (solid curve (a)).
The value of the fitted damping parameter (0.0135) is typical of
the one obtained on a NiFe single layer \cite{oliv}. For the
exchange biased bilayers $(H_{ex}^{meas}=950 A/m, H_{K}^{meas}=600
A/m, )$, the calculated resonance frequency is lower than the
measured one (Fig.\ref{figure3} solid curve (b)). In a second step
we have computed the complex permeability where the effective
field was taken as a fit parameter ($H_{eff}^{fit}$). The result
is presented in figure \ref{figure3} (solid curve (c)). The best
fit was obtained with values $H_{eff}^{fit}$=4200 A/m and
$\alpha$=0.03. It can be seen that the experimental results are in
good agreement with theoretical predictions. The fitted value of
the effective field in the exchange biased bilayers is three times
higher that the one obtained from M-H loop measurements. These
discrepancies may be due to the fact that M-H-loops involve large
applied static fields and the AC susceptibility involves only a
small perturbation of the magnetization in is "natural"
environment.
Moreover, one can see that the broadening of the experimental
spectrum of $\mu''$ is associated with the increasing of the
damping parameter from 0.012 up to 0.03.  These broadenings have
been observed with FMR and BLS measurements and attributed to a
relaxation mechanism based on two-magnon scattering processes due
to the local fluctuation of the exchange coupling caused by
interface roughness \cite{rezende1}.

\begin{figure}
\begin{center}
\includegraphics[width=7.5 cm, height=10 cm]{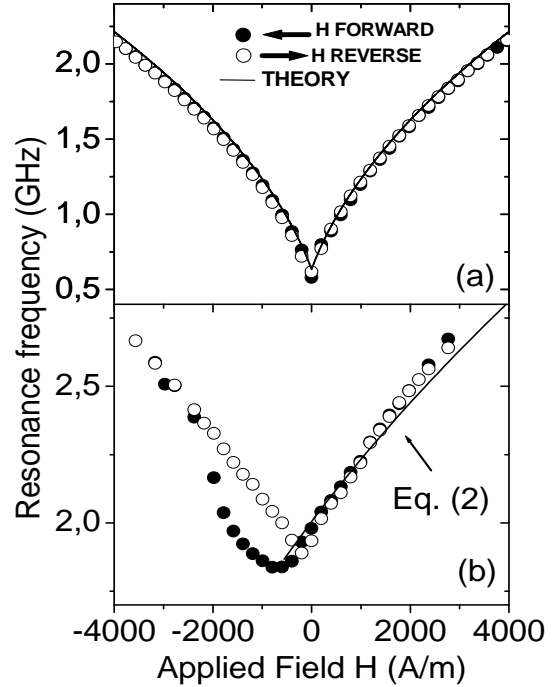}
\caption{Resonance frequencies of $\mu''$ as a function of the
applied field of a NiFe 52nm/$Mn_{50}Ni_{50}$ 80nm bilayer. (a)
as-grown; (b) annealed; full circles: forward applied field, open
circles: reverse field } \label{figure4}
\end{center}
\end{figure}

We have measured the CPS of the bilayers when submitted to a
static magnetic increasing and decreasing field applied along the
easy axis as in a hysteresis loop measurement.  These measurement
have been performed on both as-sputt and annealed bilayers, the
results are presented on Fig.\ref{figure4}. For the as-sputt
sample (Fig.\ref{figure4}(a)), the resonance mode frequency as a
function of the magnetic field is fully reversible, with a minimum
for H = 0, corresponding to the resonance frequency of 700 Mhz
mentioned above. The frequency fit well by the equation (2), where
$H_{eff}$=$H_{k}+ H$. $H_{k}$ and $M_{s}$ are taken from static
measurement. For the annealed exchange biased sample, the curves
split into two parts corresponding to the foward and reverse loop
of the hysteresis curve, the frequencies are likewise hysteretic.
This may be associated with the strong asymmetry observed in the
hysteresis loop as presented in Fig.\ref{figure1}. In the first
magnetization reversal (foward), the frequency decreases then
increases showing a minimum of 1.83 GHz for an applied field of
about -800 A/m. This minimum is not well pronounced. In the recoil
loop the frequency shows a sharp minimum of 1.89 GHz at -197 A/m.
These minima correspond roughly to the different coercive fields
in the forward and reverse magnetization direction. The evolution
of the magnetization with the forward field could indicate a
mechanism of coherent rotation, which would give a component of
the magnetization along the hard axis, which is along the applied
ac field. The may explain the decreased value of $f_{res}$ for the
forward field. The frequency on the right side of the dip is fit
very well by equation (2), where $H_{eff}$ is taken from the
effective field obtain from the previous calculation using the LL
theory and the applied static field (i.e. 4200 A/m).
 This hysteretic behavior of the resonance
frequency mode as been recently theoretically investicated by
Stamps \cite{stamps}. The author attributes this behavior to
different effectives fields governing the frequencies of resonance
for the forward and reverse field directions and obtains
hysteretic resonance frequencies by including, in the energy,
terms that take into account the coupling to both sublattices of
the antiferromagnet and the formation of a planar wall.

\section{CONCLUSION}

We have shown that it is possible to describe the magnetization
dynamics of exchange biased bilayers with the LL theory.
Effectives fields are extracted from complex permeability spectra
and are much larger with the ones obtained from hysteretis loop
measurements. The high values of the effective field are
associated with the enhancement of the damping parameter. We have
experimentally shown that in F/AF bilayers the resonance
frequencies are likewise hysteretic when the bilayers are
submitted to an forward and reverse static magnetic field. This
behavior may be related to the strong asymmetry in the hysteresis
loop.

This work was partially supported by PRIR program of Region
Bretagne.

\bibliographystyle{prsty}
\bibliography{referenc}


%
%

%
%

\end{document}